\documentclass{llncs}
\usepackage{graphicx}
\usepackage{amsmath}
\usepackage{amsfonts}
\usepackage{subcaption}
\usepackage{multirow}

\begin{document}

\author{Haixuan Guo\inst{1} \and
Shuhan Yuan \inst{1} \and
Xintao Wu \inst{2}}
\institute{Utah State University, Logan UT  \\
\email{ghaixan95@aggiemail.usu.edu, shuhan.yuan@usu.edu}\\
 \and
University of Arkansas, Fayetteville AR \\
\email{xintaowu@uark.edu}}

\title{LogBERT: Log Anomaly Detection via BERT}

\maketitle

\begin{abstract}
    Detecting anomalous events in online computer systems is crucial to protect the systems from malicious attacks or malfunctions. System logs, which record detailed information of computational events, are widely used for system status analysis. In this paper, we propose LogBERT, a self-supervised framework for log anomaly detection based on Bidirectional Encoder Representations from Transformers (BERT). LogBERT learns the patterns of normal log sequences by two novel self-supervised training tasks and is able to detect anomalies where the underlying patterns deviate from normal log sequences. The experimental results on three log datasets show that LogBERT outperforms state-of-the-art approaches for anomaly detection.
\end{abstract}

\section{Introduction}
Online computer systems are vulnerable to various malicious attacks in cyberspace. 
Detecting anomalous events from online computer systems in a timely manner is the fundamental step to protect the systems. System logs, which record detailed information about computational events generated by computer systems, play an important role in anomaly detection nowadays.

Currently, many traditional machine learning models are proposed for identifying anomalous events from log messages. These approaches extract useful features from log messages and adopt machine learning algorithms to analyze the log data. Due to the data imbalance issue, it is infeasible to train a binary classifier to detect anomalous log sequences. As a result, many unsupervised learning models, such as Principal Component Analysis (PCA) \cite{xuDetectingLargescaleSystem2009}, or one class classification models, such as one-class SVM \cite{liImprovingOneclassSVM2003,wangAnomalyIntrusionDetection2004}, are widely-used to detect anomalies. However, traditional machine learning models, such as one-class SVM, are hard to capture the temporal information of discrete log messages.

Recently, deep learning models, especially recurrent neural networks (RNNs), are widely used for log anomaly detection since they are able to model the sequential data \cite{duDeepLogAnomalyDetection2017,zhouLogAnomalyUnsupervisedDetection2019,wangMultiScaleOneClassRecurrent2021}. However, there are still some limitations of using RNN for modeling log data. 
First, although RNN can capture the sequential information by the recurrence formula, it cannot make each log in a sequence encoding the context information from both the left and right context. However, it is crucial to observe the complete context information instead of only the information from previous steps when detecting malicious attacks based on log messages. 
Second, current RNN-based anomaly detection models are trained to capture the patterns of normal sequences by prediction the next log message given previous log messages. This training objective mainly focuses on capturing the correlation among the log messages in normal sequences. When such correlation in a log sequence is violated, the RNN model cannot correctly predict the next log message based on previous ones. Then, we will label the sequence as anomalous. However, only using the prediction of next log message as objective function cannot not explicitly encode the common patterns shared by all normal sequences.

To tackle the existing limitations of RNN-based models, in this work, we propose LogBERT, a self-supervised framework for log anomaly detection based on Bidirectional Encoder Representations from Transformers (BERT). Inspired by the great success of BERT in modeling sequential text data \cite{devlinBERTPretrainingDeep2018}, we leverage BERT to capture patterns of normal log sequences. By using the structure of BERT, we expect the contextual embedding of each log entry can capture the information of whole log sequences. To achieve that, we propose two self-supervised training tasks: 1) masked log key prediction, which aims to correctly predict log keys in normal log sequences that are randomly masked; 2) volume of hypersphere minimization, which aims to make the normal log sequences close to each other in the embedding space. After training, we expect LogBERT encodes the information about normal log sequences. We then derive a criterion to detect anomalous log sequences based on LogBERT. Experimental results on three log datasets show that LogBERT achieves the best performance on log anomaly detection by comparing with various state-of-the-art baselines.

\section{Related Work}
System logs are widely used by large online computer systems for troubleshooting, where each log message is usually a semi-structured text string. The traditional approaches explicitly use the keywords (e.g., ``fail'') or regular expressions to detect anomalous log entries. However, these approaches cannot detect malicious attacks based on a sequence of operations, where each log entry looks normal, but the whole sequence is anomalous. To tackle this challenge, many rule-based approaches are proposed to identify anomalous events \cite{pecchiaEventLogsAnalysis2013,yen2013beehive}. Although rule based approaches can achieve high accuracy, they can only identify pre-defined anomalous scenarios and require heavy manual engineering. 

As malicious attacks become more complicated, learning-based approaches are proposed. The typical pipeline for these approaches consists of three steps \cite{heExperienceReportSystem2016}. First, a log parser is adopted to transform log messages to log keys. A feature extraction approach, such as TF-IDF, is then used to build a feature vector to represent a sequence of log keys in a sliding window. Finally, in most cases, an unsupervised approach is applied for detecting the anomalous sequences \cite{xu2009online,louMiningInvariantsConsole2010}. 

Recently, many deep learning-based log anomaly detection approaches are proposed for log anomaly detection \cite{zhangAutomatedITSystem2016,duDeepLogAnomalyDetection2017,zhangRobustLogbasedAnomaly2019,zhouLogAnomalyUnsupervisedDetection2019,liuLog2vecHeterogeneousGraph2019,wangMultiScaleOneClassRecurrent2021}. Most of the existing approaches adopt recurrent neural networks, especially long-short term memory (LSTM) or gated recurrent unit (GRU) to model the normal log key sequences and derive anomalous scores to detect the anomalous log sequences \cite{duDeepLogAnomalyDetection2017,zhouLogAnomalyUnsupervisedDetection2019,wangMultiScaleOneClassRecurrent2021}. In this work, we explore the advanced BERT model to capture the information of log sequences and propose two novel self-supervised tasks to train the model.

\section{LogBERT}
\begin{figure}[ht]
    \centering
    \includegraphics[width=0.95\textwidth]{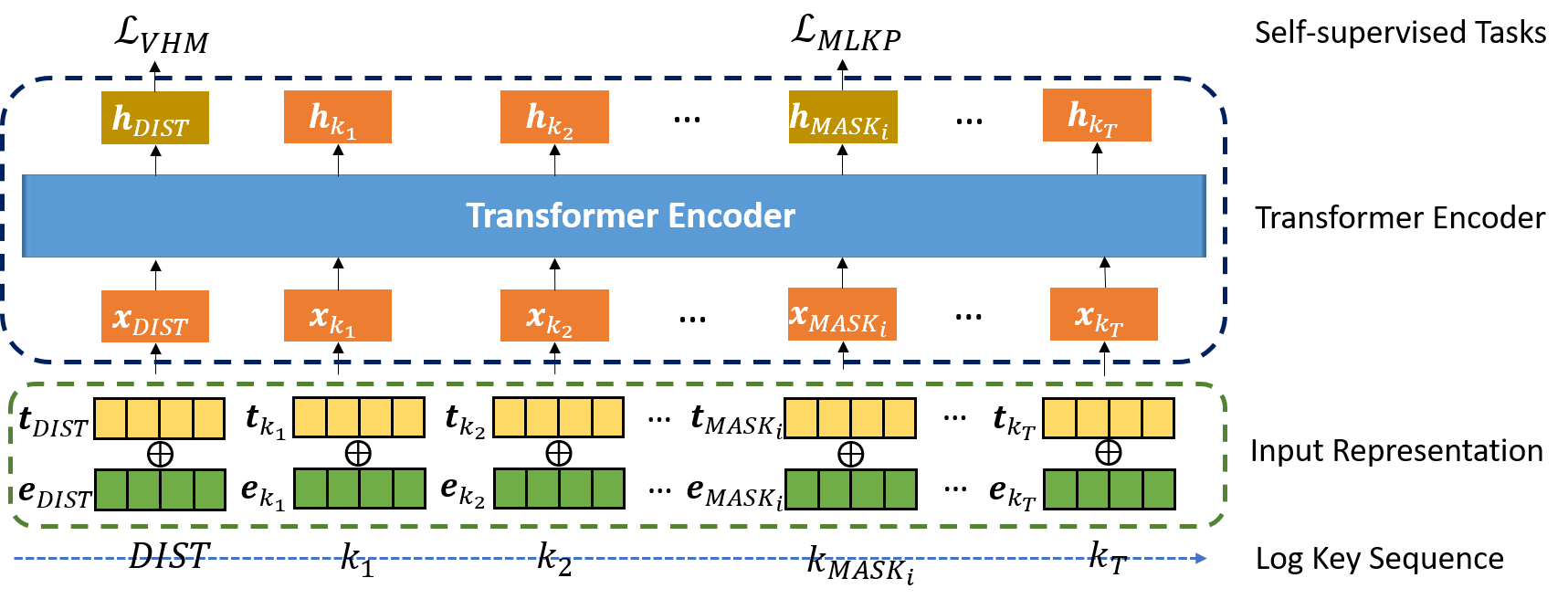}
    \caption{The overview of LogBERT}
    \label{fig:logbert}
\end{figure}

In this section, we introduce our framework, LogBERT, for log sequence anomaly detection. Inspired by BERT \cite{devlinBERTPretrainingDeep2018}, LogBERT leverages the Transformer encoder to model log sequences and is trained by novel self-supervised tasks to capture the patterns of normal sequences. Figure \ref{fig:logbert} shows the whole framework of LogBERT. 

\subsection{Framework}
\begin{figure}[ht]
    \centering
    \includegraphics[width=0.95\textwidth]{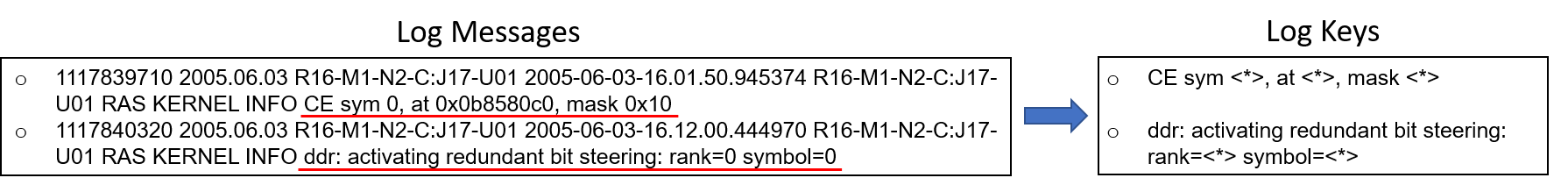}
    \caption{Log messages in the BGL dataset and the corresponding log keys extracted by a log parser. The message with red underscore indicates the detailed computational event.}
    \label{fig:logkeys}
\end{figure}

Given a sequence of unstructured log messages, we aim to detect whether this sequence is normal or anomalous. In order to represent log messages, following a typical pre-processing approach, we first extract log keys (string templates) from log messages via a log parser (shown in Figure \ref{fig:logkeys}). Then, we can define a log sequence as a sequence of ordered log keys $S=\{k_1, ..., k_t, ..., k_T\}$, where $k_t \in \mathcal{K}$ indicates the log key in the $t$-th position, and $\mathcal{K}$ indicates a set of log keys extracted from log messages. The goal of this task is to predict whether a new log sequence $S$ is anomalous based on a training dataset $\mathcal{D}=\{S^j\}_{j=1}^N$ that consists of only normal log sequences. To achieve that, LogBERT models the normal sequences and further derive an anomaly detection criterion to identify anomalous sequences. 

{\bf \noindent Input Representation.}
Given a normal log sequence $S^j$, we first add a special token, DIST, at the beginning of $S^j=\{k^j_1, ..., k^j_t, ..., k^j_T\}$ as the first log key, which is used to represent the whole log sequence based on the structure of Transformer encoder. LogBERT then represents each log key $k^j_t$ as an input representation $\mathbf{x}^j_t$, where the representation $\mathbf{x}^j_t$ is a summation of a log key embedding and a position embedding.  In this work, we randomly generate a matrix $\mathbf{E} \in \mathbb{R}^{|\mathcal{K}|*d}$ as the log key embedding matrix, where $d$ is the dimension of log key embedding, while the position embeddings $\mathbf{T} \in \mathbb{R}^{T*d}$ are generated by using a sinusoid function to encode the position information of log keys in a sequence \cite{devlinBERTPretrainingDeep2018}. Finally, the input representation of the log key $k_t$ is defined as: $\mathbf{x}^j_t=\mathbf{e}_{k^j_t}+\mathbf{t}_{k^j_t}$.

{\bf \noindent Transformer Encoder.}
LogBERT adopts Transformer encoder to learn the contextual relations among log keys in a sequence. Transformer encoder consists of multiple transformer layers. Each transformer layer includes a multi-head self-attention and a position-wise feed forward sub-layer in which a residual connection is employed around each of two sub-layers, followed by layer normalization \cite{vaswaniAttentionAllYou2017}. The multi-head attention employs $H$ parallel self-attentions to jointly capture different aspect information at different positions over the input log sequence. 
Formally, for the $l$-th head of the attention layer, the scaled dot-product self-attention is defined as:
\begin{equation}
  head_l = Attention(\mathbf{X}^j\mathbf{W}_l^Q, \mathbf{X}^j\mathbf{W}_l^K, \mathbf{X}^j\mathbf{W}_l^V),
\end{equation}
where $Attention(\mathbf{Q}, \mathbf{K},\mathbf{V})=softmax(\frac{\mathbf{Q}\mathbf{K}^T}{\sqrt{d_v}})\mathbf{V}$; $\mathbf{X}^j \in \mathbb{R}^{T*d}$ is the input representation of the log sequence; $\mathbf{W}_l^Q$, $\mathbf{W}_l^K$ and $\mathbf{W}_l^V$ are linear projection weights with dimensions $\mathbb{R}^{d*d_v}$ for the $l$-th head, and $d_v$ is the dimension for one head ot the attention layer. Each self-attention makes each key attend to all the log keys in an input sequence and computes the hidden representation for each log key with an attention distribution over the sequence. 

The multi-head attention employs a parallel of self-attentions to jointly capture different aspect information at different log keys. Formally, the multi-head attention concatenates $H$ parallel heads together as:
\begin{equation}
  f(\mathbf{X}^j) = Concat(head_1,...,head_H)\mathbf{W}^O,
\end{equation}
where $\mathbf{W}^O \in \mathbb{R}^{{hd_v}*d_o}$ is a projection matrix, and $d_o$ is the dimension for the output of multi-head attention sub-layer. 

Then, the position-wise feed forward sub-layer with a ReLU activation is applied to the hidden representation of each activity separately. Finally, by combining the position-wise feed forward sub-layer and multi-head attention, a transformer layer is defined as:
\begin{equation}
% \begin{split}
    \text{transformer\_layer}(\mathbf{X}^j) = FFN(f(\mathbf{X}^j))
    = ReLU(f(\mathbf{X}^j) \mathbf{W}_1)\mathbf{W}_2,
% \end{split}
\end{equation}
where $\mathbf{W}_1$ and $\mathbf{W}_2$ are trained projection matrices. 

The Transformer encoder usually consists of multiple transformer layers. We denote $\mathbf{h}^j_t$ as the contextual embedding vector of the log key $k^j_t$ produced by the Transformer encoder, i.e., $\mathbf{h}^j_t = \text{Transformer}(x^j_t)$.

\subsection{Objective Function}
In order to train the LogBERT model, we propose two self-supervised training tasks to capture the patterns of normal log sequences.

{\bf \noindent Task I: Masked Log Key Prediction (MLKP).}
In order to capture the bidirectional context information of log sequences, we train LogBERT to predict the masked log keys in log sequences. In our scenario, LogBERT takes log sequences with random masks as inputs, where we randomly replace a ratio of log keys in a sequence with a specific MASK token. The training objective is to accurately predict the randomly masked log keys. The purpose is to make LogBERT encode the prior knowledge of normal log sequences. 

To achieve that, we feed the contextual embedding vector of the $i$-th MASK token in the $j$-th log sequence $\mathbf{h}^j_{[\text{MASK}_i]}$ to a softmax function, which will output a probability distribution over the entire set of log keys $\mathcal{K}$:
\begin{equation}
    \centering
    \mathbf{\hat{y}}^j_{[\text{MASK}_i]} = Softmax(\mathbf{W}_C\mathbf{h}^j_{[\text{MASK}_i]}+\mathbf{b}_C),
    \label{eq:y_hat}
\end{equation}
where $\mathbf{W}_C$ and $\mathbf{b}_C$ are trainable parameters. Then, we adopt the cross entropy loss as the objective function for masked log key prediction, which is defined as:
\begin{equation}
\centering
    \mathcal{L}_{MLKP} =- \frac{1}{N}\sum^N_{j=1}{\sum^M_{i=1}{\mathbf{y}^j_{[\text{MASK}_i]}\log{\mathbf{\hat{y}}^j_{{[\text{MASK}_i]}}}}},
    \label{eq:mlkp_loss}
\end{equation}
where $\mathbf{y}^j_{[\text{MASK}_i]}$ indicates the real log key for the $i$-th masked token, and $M$ is the total number of masked tokens in the $j$-th log sequence.
Since the patterns of normal and anomalous log sequences are different, we expect once LogBERT is able to correctly predict the masked log keys, it can distinguish the normal and anomalous log sequences. 

{\bf \noindent Task II: Volume of Hypersphere Minimization (VHM).}
Inspired by the Deep SVDD approach \cite{ruffDeepOneClassClassification2018}, where the objective is to minimize the volume of a data-enclosing hypersphere, we propose a spherical objective function to regulate the distribution of normal log sequences. The motivation is that normal log sequences should be concentrated and close to each other in the embedding space, while the anomalous log sequences are far to the center of the sphere. We first derive the representations of normal log sequences and then compute the center representation based on the mean operation. In particular, we consider the contextual embedding vector of the DIST token $\mathbf{h}^j_{\text{DIST}}$, which encodes the information of entire log sequence based on the Transformer encoder, as the representation of a log sequence in the embedding space. To make the representations of normal log sequences close to each other, we further derive the center representation of normal log sequences $\mathbf{c}$ in the training set by a mean operation, i.e., $\mathbf{c}=\text{Mean}(\mathbf{h}^j_{\text{DIST}})$. Then, the objective function is to make the representation of normal log sequence  $\mathbf{h}^j_{\text{DIST}}$ close to the center representation $\mathbf{c}$:
\begin{equation}
    \centering
    \mathcal{L}_{VHM} =\frac{1}{N} \sum_{j=1}^N{||\mathbf{h}^j_{\text{DIST}}-\mathbf{c}||^2}.
    \label{eq:svdd_loss}
\end{equation}
By minimizing the Equation \ref{eq:svdd_loss}, we expect all the normal log sequences in the training set are close to the center, while the anomalous log sequences have a larger distance to the center. Meanwhile, another advantage of the spherical objective function is that by making the sequence representations close to the center, the Transformer encoder can also leverage the information from other log sequences via the center representation $\mathbf{c}$, since $\mathbf{c}$ encodes all the information of normal log sequences. As a result, the model should be able to predict the masked log keys with higher accuracy for normal log sequences because the normal log sequences should share similar patterns.

Finally, the objective function for training the LogBERT is defined as below:
\begin{equation}
    \centering
    \mathcal{L} = \mathcal{L}_{MLKP} + \alpha \mathcal{L}_{VHM},
    \label{eq:loss}
\end{equation}
where $\alpha$ is a hyper-parameter to balance two training tasks.

\subsection{Anomaly Detection}
After training, we can deploy LogBERT for anomalous log sequence detection. The idea of applying LogBERT for log anomaly detection is that since LogBERT is trained on normal log sequences, it can achieve high prediction accuracy on predicting the masked log keys if a testing log sequence is normal. Hence, we can derive the anomalous score of a log sequence based on the prediction results on the MASK tokens. 
To this end, given a testing log sequence, similar to the training process, we first randomly replace a ratio log keys with MASK tokens and use the randomly-masked log sequence as an input to LogBERT. Then, given a MASK token, the probability distribution calculated based on Equation \ref{eq:y_hat} indicates the likelihood of a log key appeared in the position of the MASK token. Similar to the strategy in DeepLog\cite{duDeepLogAnomalyDetection2017}, we build a candidate set consisting of $g$ normal log keys with the top $g$ highest likelihoods computed by $\mathbf{\hat{y}}_{[\text{MASK}_i]}$. If the real log key is in the candidate set, we treat the key as normal. In other words, if the observed log key is not in the top-$g$ candidate set predicted by LogBERT, we consider the log key as an anomalous log key. Then, when a log sequence consists of more than $r$ anomalous log keys, we will label this log sequence as anomalous. Both $g$ and $r$ are hyper-parameters and will be tuned based on a validation set.

\section{Experiments}
\subsection{Experimental Setup}
{\bf \noindent Datasets.}
We evaluate the proposed LogBERT on three log datasets, HDFS, BGL, and Thunderbird. Table \ref{tb:datasets} shows the statistics of the datasets. For all datasets, we adopt around 5000 normal log sequences for training. The number in the brackets under the column ``\# Log Keys'' indicates the number of unique log keys in the training dataset. 
\begin{table}[h]
\caption{Statistics of evaluation datasets}
\label{tb:datasets}
\resizebox{.98\textwidth}{!}{
    \begin{tabular}{|c|c|c|c|c|c|}
        \hline
        \multirow{2}{*}{Dataset} & \multirow{2}{*}{\# Log Messages} & \multirow{2}{*}{\# Anoamlies} & \multirow{2}{*}{\# Log Keys} & \multicolumn{2}{c|}{\# of Log Sequences in Test Dataset} \\ \cline{5-6} 
                                 &                                  &                               &                              & Normal                    & Anomalous                    \\ \hline
        HDFS                     & 11,172,157                       & 284,818                       & 46 (15)                         & 553,366                    & 10,647                        \\ \hline
        BGL                      & 4,747,963                        & 348,460                       & 334 (175)                        & 10,045                     & 2,630                         \\ \hline
        Thunderbird-mini         & 20,000,000                       &  758,562                      &1,165 (866)                  &  71,155                    & 45,385                            \\ \hline
        \end{tabular}
        }
\end{table}

\begin{itemize}
    \item Hadoop Distributed File System (HDFS) \cite{xu2009online}. HDFS dataset is generated by running Hadoop-based map-reduce jobs on Amazon EC2 nodes and manually labeled through handcrafted rules to identify anomalies. HDFS dataset consists of 11,172,157 log messages, of which 284,818 are anomalous. For HDFS, we group log keys into log sequences based on the session ID in each log message. The average length of log sequences is 19.

    \item BlueGene/L Supercomputer System (BGL) \cite{oliner2007supercomputers}. BGL dataset is collected from a BlueGene/L supercomputer system at Lawrence Livermore National Labs (LLNL). Logs contain alert and non-alert messages identified by alert category tags. The alert messages are considered as anomalous. BGL dataset consists of 4,747,963 log messages, of which 348,460 are anomalous. For BGL, we define a time sliding window as 5 minutes to generate log sequences, where the average length is 562.

    \item Thunderbird \cite{oliner2007supercomputers}. Thunderbird dataset is another large log dataset collected from a supercomputer system. We select the first 20,000,000 log messages from the original Thunderbird dataset to compose our dataset, of which 758,562 are anomalous. For Thunderbird, we also adopt a time sliding window as 1 minute to generate log sequences, where the average length is 326.
\end{itemize}

{\bf \noindent Baselines.}
We compare our LogBERT model with the following baselines.
\begin{itemize}
    \item Principal Component Analysis (PCA) \cite{xuDetectingLargescaleSystem2009}. PCA builds counting matrix based on the frequency of log keys sequences and then reduces the original counting matrix into a low dimensional space to detect anomalous sequences.
    \item One-Class SVM (OCSVM) \cite{scholkopfEstimatingSupportHighDimensional2001}. One-Class SVM is a well-known one-class classification model and widely used for log anomaly detection \cite{liImprovingOneclassSVM2003,wangAnomalyIntrusionDetection2004} by only observing the normal data.
    \item IsolationForest (iForest) \cite{liuIsolationForest2008}. Isolation forest is an unsupervised learning algorithm for anomaly detection by representing features as tree structures.
    \item LogCluster \cite{linLogClusteringBased2016}. LogCluster is a clustering based approach, where the anomalous log sequences are detected by comparing with the existing clusters.
    \item DeepLog \cite{duDeepLogAnomalyDetection2017}. DeepLog is a state-of-the-art log anomaly detection approach. DeepLog adopts recurrent neural network to capture patterns of normal log sequences and further identifies the anomalous log sequences based on the performance of log key predictions.
    \item LogAnomaly \cite{zhouLogAnomalyUnsupervisedDetection2019}. Log Anomaly is a deep learning-based anomaly detection approach and able to detect sequential and quantitative log anomalies. 
\end{itemize}

{\bf \noindent Implementation Details.}
We adopt Drain \cite{heDrainOnlineLog2017} to parse the log messages into log keys. Regarding baselines, we leverage the package \textit{Loglizer} \cite{heExperienceReportSystem2016} to evaluate PCA, OCSVM, iForest as well as LogCluster for anomaly detection and adopt the open source deep learning-based log analysis toolkit to evaluate DeepLog and LogAnomaly \footnote{\url{https://github.com/donglee-afar/logdeep}}. For LogBERT, we construct a Transformer encoder by using two Transformer layers. The dimensions for the input representation and hidden vectors are 50 and 256, respectively. The hyper-parameters, including $\alpha$ in Equation \ref{eq:loss}, $m$ the ratio of masked log keys for the MKLP task, $r$ the number of predicted anomalous log keys, and $g$ the size of top-$g$ candidate set for anomaly detection are tuned based on a small validation set. In our experiments, both training and detection phases have the same ratio of masked log keys $m$. The code of LogBERT is available online \footnote{\url{https://github.com/HelenGuohx/logbert}}.

\subsection{Experimental Results}

\begin{table}[]
    \centering
    \caption{Experimental Results on HDFS, BGL, and Thunderbird Datasets}
    \label{tb:results}
    \resizebox{.98\textwidth}{!}{
    \begin{tabular}{|c|c|c|c|c|c|c|c|c|c|}
    \hline
    \multirow{2}{*}{Method} & \multicolumn{3}{c|}{HDFS}                         & \multicolumn{3}{c|}{BGL}                         & \multicolumn{3}{c|}{Thunderbird}                  \\ \cline{2-10} 
                            & Precision      & Recall          & F-1 score      & Precision      & Recall         & F-1 score      & Precision      & Recall          & F-1 score      \\ \hline
    PCA                     & 5.89           & 100.00 & 11.12          & 9.07           & 98.23 & 16.61          & 37.35          & 100.00 & 54.39          \\ \hline
    iForest         & 53.60          & 69.41           & 60.49          & 99.70 & 18.11          & 30.65          & 34.45          & 1.68            & 3.20           \\ \hline
    OCSVM                   & 2.54           & 100.00 & 4.95           & 1.06           & 12.24          & 1.96           & 18.89          & 39.11           & 25.48          \\ \hline
    LogCluster           & 99.26 & 37.08           & 53.99          & 95.46          & 64.01          & 76.63          & 98.28 & 42.78           & 59.61          \\ \hline
    DeepLog                 & 88.44          & 69.49           & 77.34          & 89.74          & 82.78          & 86.12          & 87.34          & 99.61           & 93.08          \\ \hline
    LogAnomaly              & 94.15          & 40.47           & 56.19          & 73.12          & 76.09          & 74.08          & 86.72          & 99.63           & 92.73          \\ \hline
    LogBERT                 & 87.02          & 78.10           & \textbf{82.32} & 89.40          & 92.32          & \textbf{90.83} & 96.75          & 96.52           & \textbf{96.64} \\ \hline
    \end{tabular}
    }
    \end{table}

{\bf \noindent Performance on Log Anomaly Detection.}
Table \ref{tb:results} shows the results of LogBERT as well as baselines on three datasets. We can notice that PCA, Isolation Forest, and OCSVM have poor performance on log anomaly detection. Although these methods could achieve extremely high precision or recall values, they cannot balance the performance on both precision and recall, which lead to extremely low F1 scores. This could be because using the counting vector to represent a log sequence leads to the loss of temporal information from sequences. LogCluster, which is designed for log anomaly detection, achieves better performance than the PCA, Isolation Forest, and OCSVM. Meanwhile, two deep learning-based baselines, DeepLog and LogAnomaly, significantly outperform the traditional approaches and achieve reasonable F1 scores on three datasets, which show the advantage to adopt deep learning models to capture the patterns of log sequences. Moreover, our proposed LogBERT achieves the highest F1 scores on three datasets with large margins by comparing with all baselines. It indicates that by using self-supervised training tasks, LogBERT can successfully model the normal log sequences and further identify anomalous sequences with high accuracy.

\begin{table}[]
    \caption{Performance of LogBERT base on One Self-supervised Training Task}
    \label{tb:ablation}
    \resizebox{.98\textwidth}{!}{
        \begin{tabular}{|c|c|c|c|c|c|c|c|c|c|}
        \hline
        \multirow{2}{*}{} & \multicolumn{3}{c|}{HDFS}      & \multicolumn{3}{c|}{BGL}       & \multicolumn{3}{c|}{Thunderbird} \\ \cline{2-10} 
                                & Precision & Recall & F-1 score & Precision & Recall & F-1 score & Precision  & Recall  & F-1 score \\ \hline
        MLKP                & 77.54     & 78.65  & 78.09     & 93.16     & 86.46  & 89.69     & 97.07      & 95.90   & 96.48     \\ \hline
        VHM              & 2.43      & 39.17  & 4.58      & 71.04     & 43.84  & 54.22    & 56.58      & 43.87   & 49.42     \\ \hline
        Both                 & 87.02     & 78.10  & 82.32     & 89.40     & 92.32  & 90.83     & 96.75      & 96.52   & 96.64     \\ \hline
        \end{tabular}
    }
    \end{table}
{\bf \noindent Ablation Studies.}
In order to further understand our proposed LogBERT, we conduct ablation experiments on three log datasets. LogBERT is trained by two self-supervised tasks. We evaluate the performance of LogBERT by only using one training task each time. When the model is only trained by minimizing the volume of hypersphere, we identify anomalous log sequences by computing distances of the log sequence representations to the center of normal log sequences $\mathbf{c}$. If the distance is larger than a threshold, we consider a log sequence is anomalous. 
Table \ref{tb:ablation} shows the experimental results. We can notice that when only using the task of masked log key prediction to train the model, we can still get very good performance on log anomaly detection, which shows the effectiveness of training the model by predicting masked log keys. We can also notice that even we do not train the LogBERT with the task of the volume of hypersphere minimization, LogBERT achieves higher F1 scores than DeepLog on all three datasets, which shows that compared with LSTM, Transformer encoder is better at capturing the patterns of log sequences. Meanwhile, we can observe that when only training the model for minimizing the volume of hypersphere, the performance is poor. It indicates that only using distance as a measure to identify anomalous log sequences cannot achieve good performance. However, combining two self-supervised tasks to train LogBERT can achieve better performance than the models only trained by one task. Especially, for the HDFS dataset, LogBERT trained by two self-supervised tasks gains a large margin in terms of F1 score (82.32) compared with the model only trained by MLKP (78.09). For BGL and Thunderbird, the improvement of LogBERT is not as significant as the model in HDFS. This could be because the average length of log sequences in BGL (562) and Thunderbird (326) datasets are much larger than the log sequences in HDFS (19). For longer sequences, only predicting the masked log keys can capture the most important patterns of log sequences since there are many more mask tokens in longer sequences. On the other hand, for short log sequences, we cannot have many masks tokens. As a result, the task of the volume of hypersphere minimization can help to boost the performance. 
Hence, based on Table \ref{tb:ablation}, we can conclude that using two self-supervised tasks to train LogBERT can achieve better performance, especially when the log sequences are relatively short.

\begin{figure}
    \centering
    \begin{subfigure}[b]{0.45\textwidth}
        \centering
        \includegraphics[width=\textwidth]{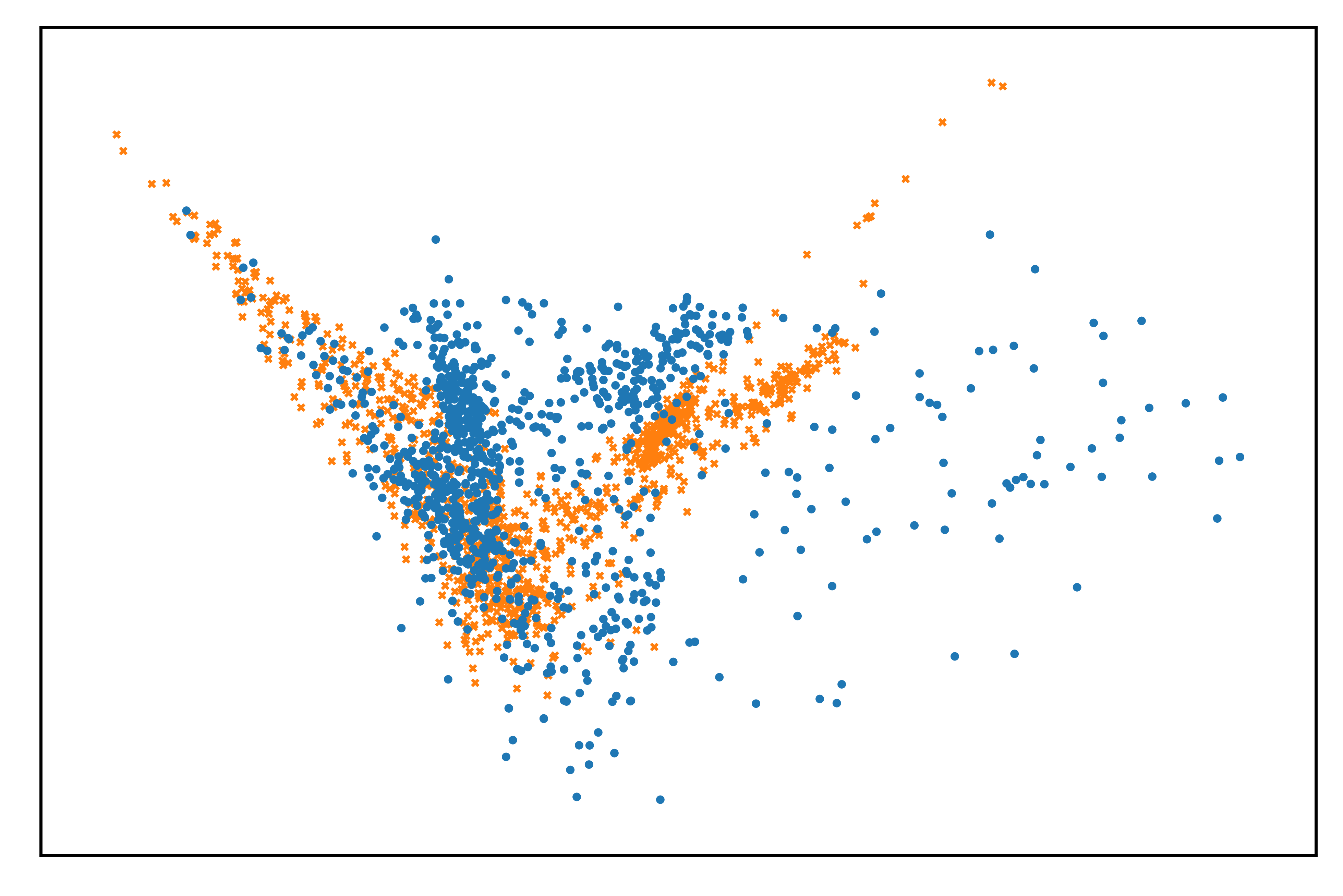}
        \caption{Trained without the VHM task}
        \label{fig:without_vhm}
      \end{subfigure}
      \hfill
      \begin{subfigure}[b]{0.45\textwidth}
        \centering
        \includegraphics[width=\textwidth]{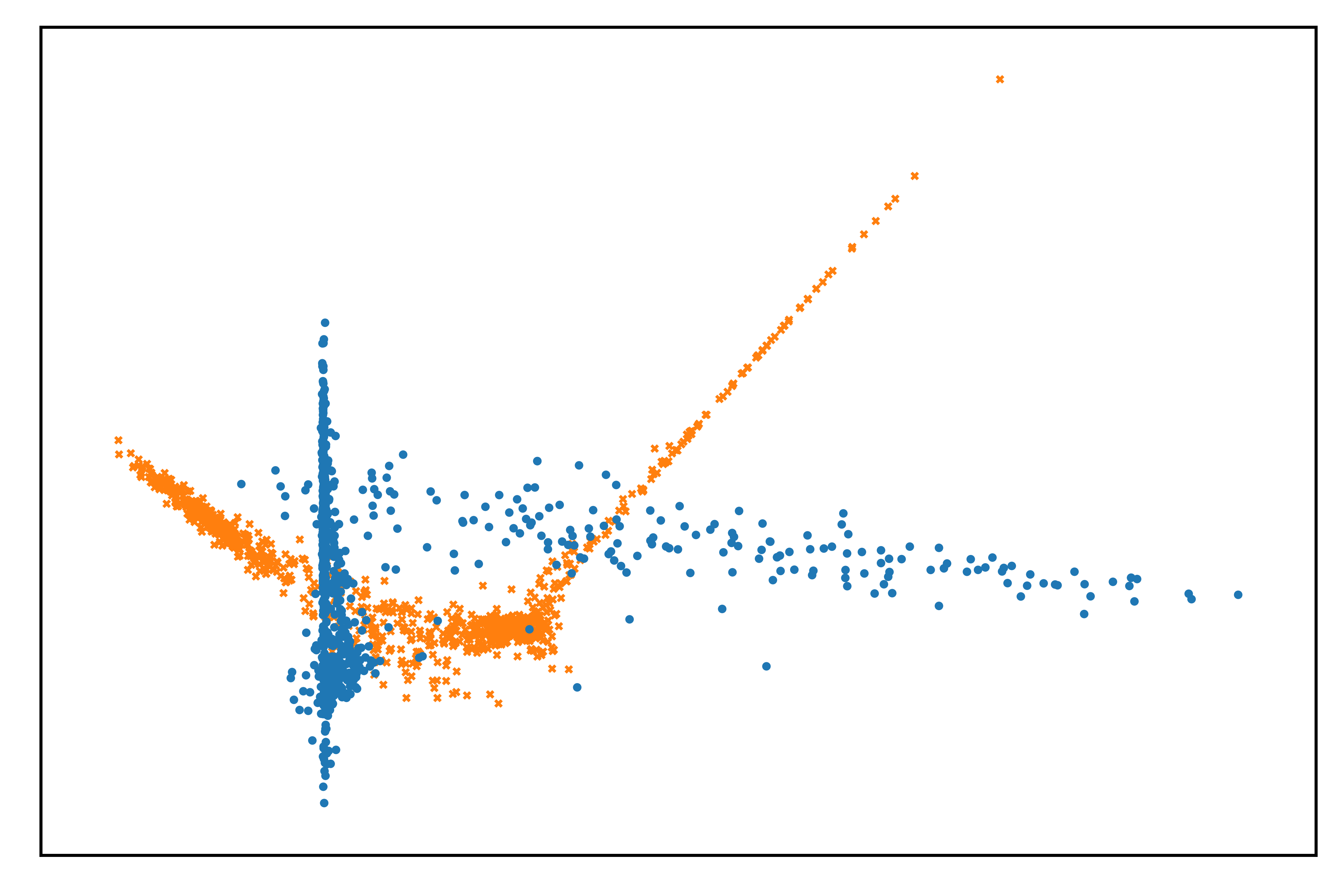}
        \caption{Trained by MLKP and VHM tasks}
        \label{fig:with_vhm}
      \end{subfigure}
    \caption{Visualization of log sequences by using the contextual embedding of DIST tokens $\mathbf{h}_{\text{DIST}}$. The blue dots indicate the normal log sequences, while the orange `x' symbols indicate anomalous log sequences.}
    \label{fig:vis}
\end{figure}
{\bf \noindent Visualization.}
In order to visualize the log sequences, we adopt locally linear embedding (LLE) algorithm \cite{roweisNonlinearDimensionalityReduction2000} to map the log sequence representations into a two dimensional space, where the contextual embedding of DIST token $\mathbf{h}_{\text{DIST}}$ is used as the representation of a log sequence. We randomly select 1000 normal and 1000 anomalous sequences from the HDFS dataset for visualization. Figure \ref{fig:vis} shows the visualization results of log sequences trained by LogBERT with and without the VHM task. We can notice that the normal and anomalous log sequences are mixed together when we trained the model without the VHM task (shown in Figure \ref{fig:without_vhm}). On the contrary, as shown in Figure \ref{fig:with_vhm}, by incorporating the VHM task, the normal and anomalous log sequences are clearly separated in the latent space, and the normal log sequences group together. Therefore, the visualization presents that the VHM task is effective in regulating the model to split the normal and abnormal data in latent space.

\begin{figure}
    \centering
    \begin{subfigure}[b]{0.32\textwidth}
        \centering
        \includegraphics[width=\textwidth]{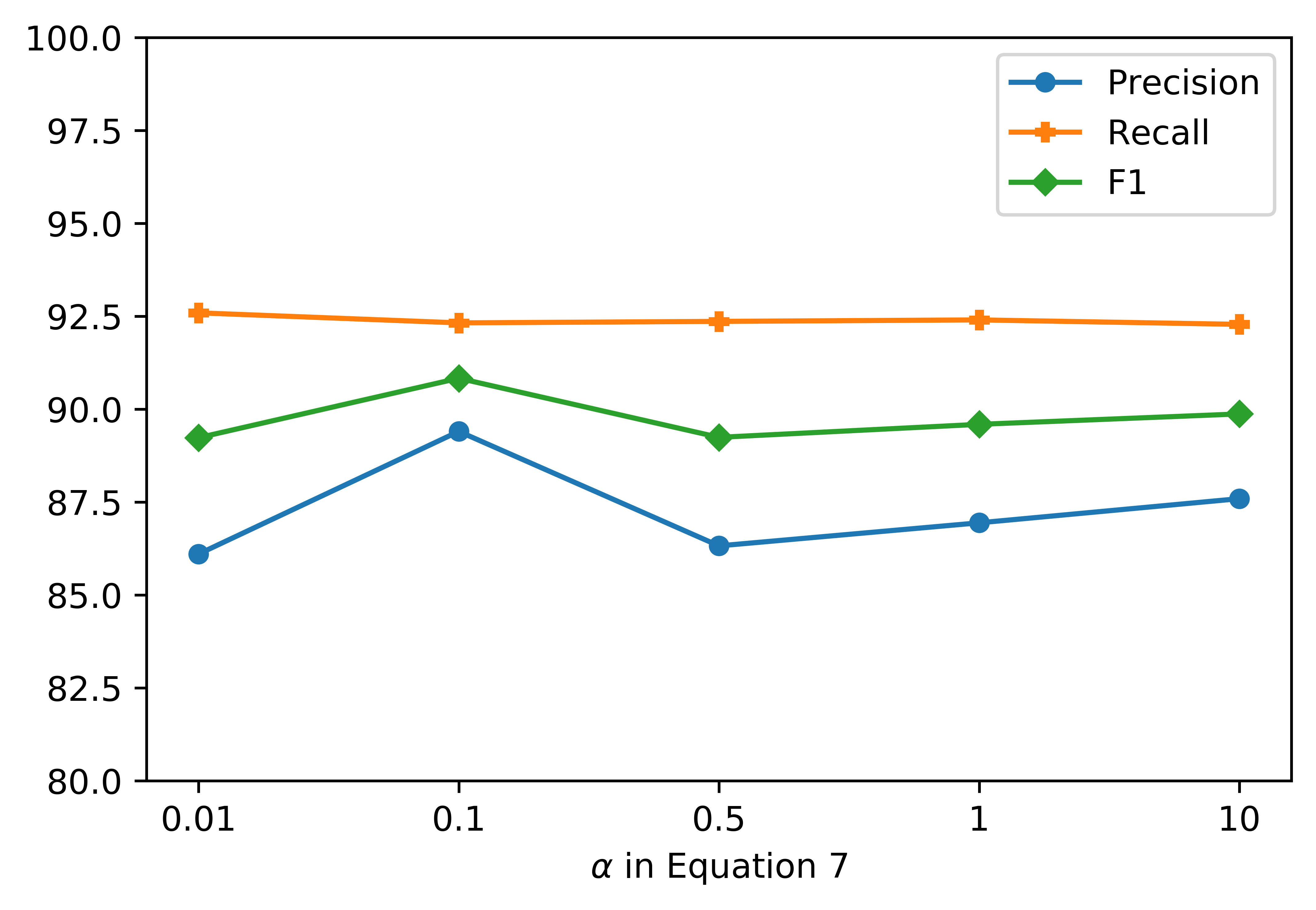}
        \caption{Different $\alpha$ values}
        \label{fig:alpha}
      \end{subfigure}
      \hfill
      \begin{subfigure}[b]{0.32\textwidth}
        \centering
        \includegraphics[width=\textwidth]{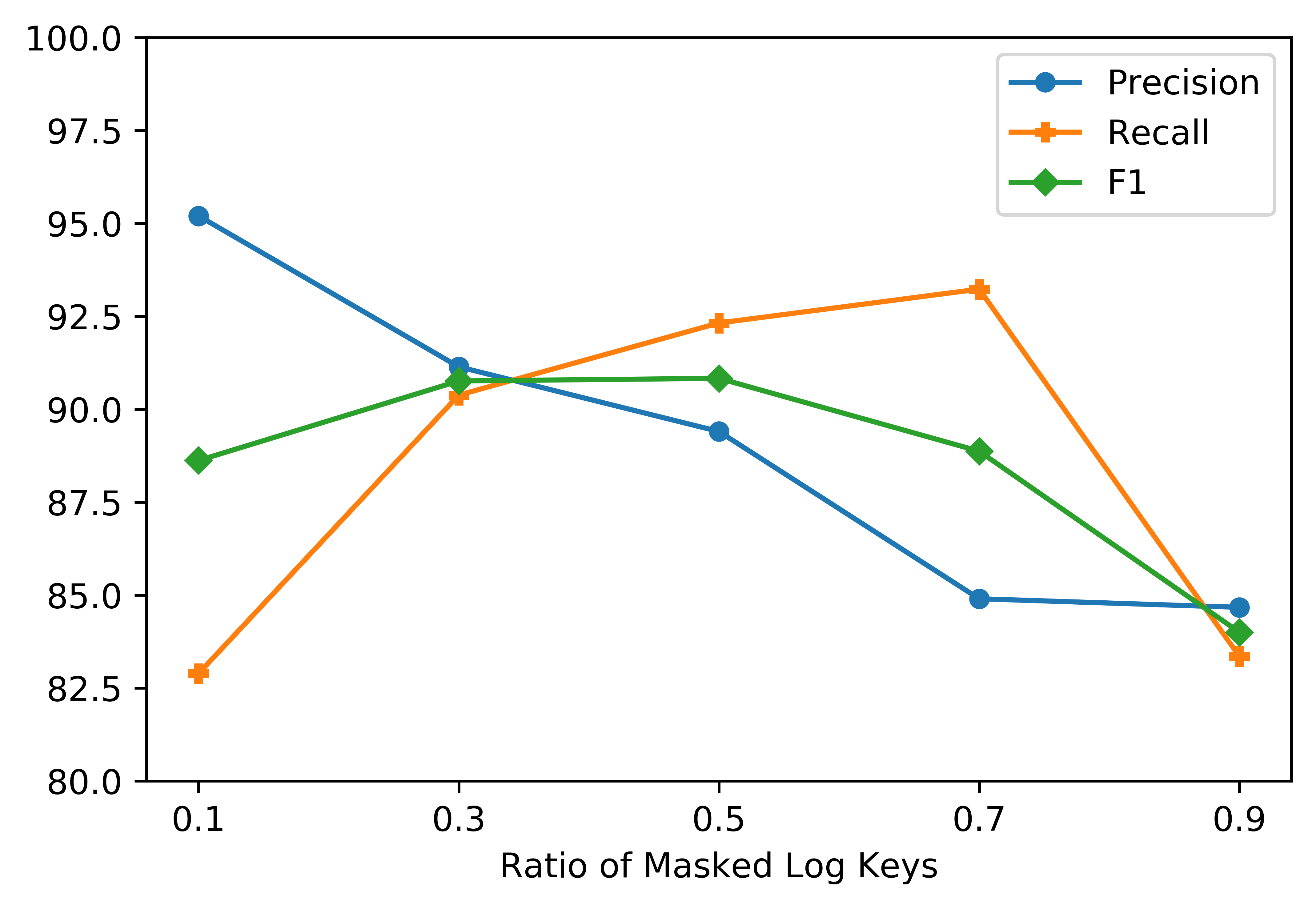}
        \caption{Different ratios of masks}
        \label{fig:mask_ratio}
      \end{subfigure}
      \hfill
      \begin{subfigure}[b]{0.32\textwidth}
        \centering
        \includegraphics[width=\textwidth]{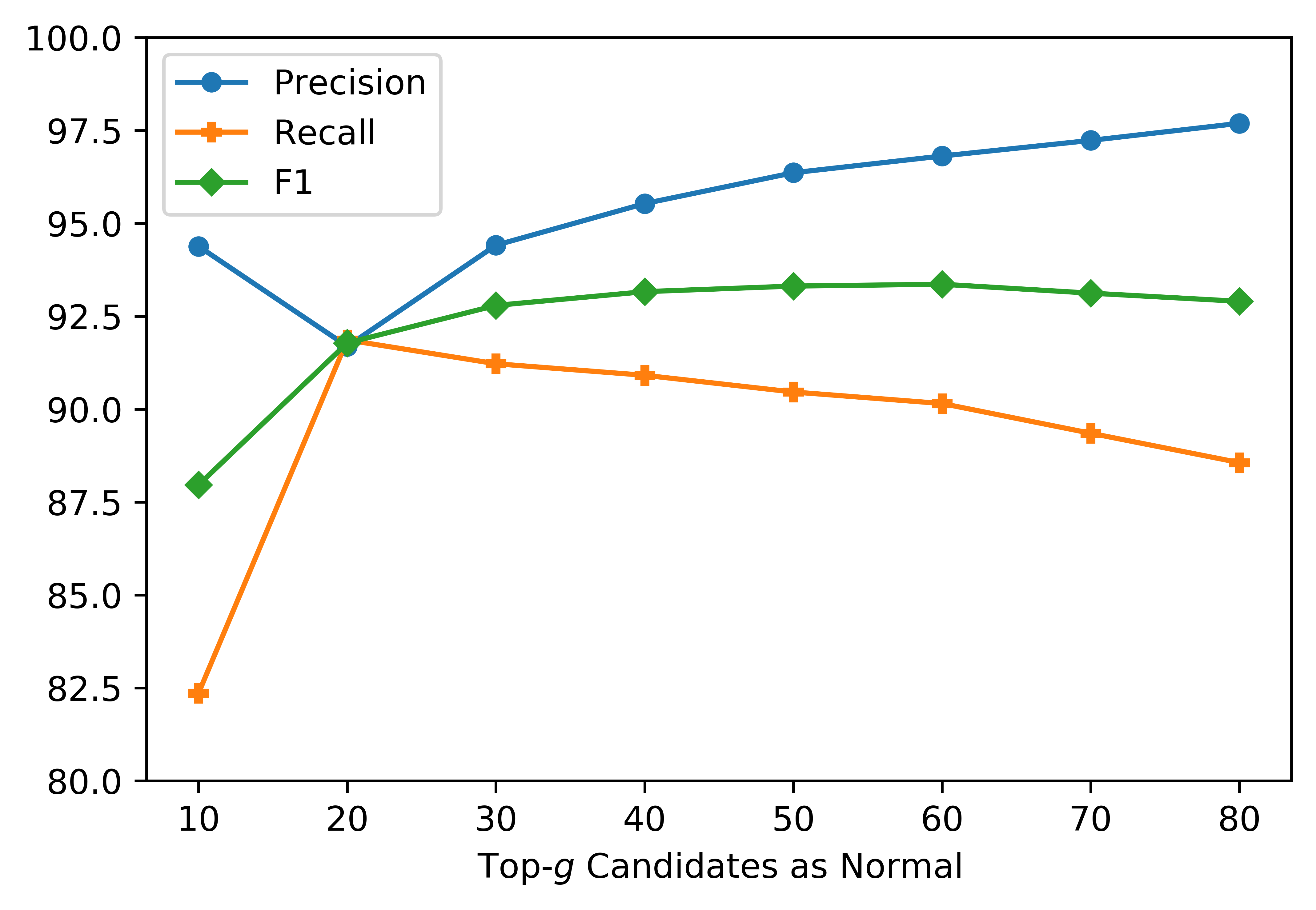}
        \caption{Different $g$-candidates}
        \label{fig:g_candidates}
      \end{subfigure}
    \caption{Parameter analysis on the BGL dataset.}
    \label{fig:parameter}
\end{figure}
{\bf \noindent Parameter analysis.}
We analyze the sensitivity of model performance by tuning various hyper-parameters. Figure \ref{fig:alpha} shows that the model performance is relatively stable by setting different $\alpha$ values in Equation \ref{eq:loss}. This is because, for the BGL dataset, the loss from the masked log key prediction dominates the final loss value due to the long log sequences. As a result, the weight for the VHM task does not have much influence on the performance. 
Figure \ref{fig:mask_ratio} shows the performance with different ratios of masked log keys. Note that we use the same ratio in both training and detection phases. We can notice that increasing the ratios of masked log keys in the sequences from 0.1 to 0.5 can slightly increase the F1 scores while keeping increasing the ratios makes the performance worse. This is because while the masked log keys increase in a reasonable range, the model can capture more information about the sequence. However, if a sequence contains too many masked log keys, it loses too much information for making the predictions. 
Figure \ref{fig:g_candidates} shows that when increasing the size of the candidate set as normal log keys, the precision for anomaly detection keeps increasing while the recall is reducing, which meets our expectation. Hence, we need to find the appropriate size of the candidate set to balance the precision and recall for the anomaly detection. 

\section{Conclusion}
Log anomaly detection is essential to protect online computer systems from malicious attacks or malfunctions. In this paper, we have developed LogBERT, a novel log anomaly detection model based on BERT. In order to train LogBERT only based on normal log sequences, we have proposed two self-supervised training tasks. One is to predict the masked log keys in log sequences, while the other is to make the normal log sequences close to each other in the embedding space. After training over normal log sequences, LogBERT is able to detect anomalous log sequences. Experimental results on three log datasets have shown that LogBERT outperforms the state-of-the-art approaches for log anomaly detection.

\bibliographystyle{splncs04}

\end{document}